\documentclass[aps,prd,reprint,onecolumn,notitlepage,superscriptaddress]{revtex4-1}

\usepackage{lipsum}
\usepackage{amsmath}
\usepackage{graphicx}
\usepackage{color}
\usepackage{tikz}
\usepackage{verbatim}
\usepackage{amssymb}
\usepackage{bm}
\newcommand{\rom}[1]{\textup{\uppercase\expandafter{\romannumeral#1}}}

\begin{document}

\title{Higher Derivative Scalar Tensor Theory in Unitary Gauge}
\author{Pawan Joshi\footnote{email:pawanjoshi697@iiserb.ac.in}
       }

\affiliation{Indian Institute of Science Education and Research Bhopal,\\ Bhopal 462066, India}

\author{Sukanta Panda \footnote{email:sukanta@iiserb.ac.in}
       }

\affiliation{Indian Institute of Science Education and Research Bhopal,\\ Bhopal 462066, India}
\begin{abstract}
Ostrogradsky instability generally appears in nondegenerate higher-order derivative theories and this issue can be resolved by removing any existing degeneracy present in such theories. We consider an action involving terms that are at most quadratic in second derivatives of the scalar field and non-minimally coupled with the curvature tensors. We perform a 3+1 decomposition of the Lagrangian to separate second-order time derivative terms from rest. This decomposition is useful for checking the degeneracy hidden in the Lagrangian and helps us find conditions under which Ostrogradsky instability does not appear. We show that our construction of Lagrangian resembles that of a GR-like theory for a particular case in the unitary gauge. As an example, we calculate the equation of motion for the flat FRW. We also write the action for open and closed cases, free from higher derivatives for a particular choice derived from imposing degeneracy conditions.

\end{abstract}

\maketitle

\section{Introduction}
Several studies have pointed out that there are many ways to explain the current accelerated expansion of the universe\cite{RiessAdamG:1998cb, Capozziello:2010zz, Capozziello:2011et}. The most popular approach involves the use of Einstein-Hilbert's action with a cosmological constant term, which unfortunately suffers from a fine-tuning problem that is purely theoretical in nature\cite{RevModPhys.61.1}. Another possible way is to modify the stress-energy sector by introducing dynamical dark energy\cite{PhysRevD.37.3406, Peebles:1987ek, Carroll:1998zi} or by introducing additional scalar degrees of freedom (dof) along with two tensor dof of general relativity (GR) by modifying the gravity part of the action. The latter can be realized as scalar-tensor theories. A generic scalar-tensor theory would contain higher derivative terms in the scalar field as well as in metric tensor. 

In this direction, higher derivative scalar-tensor theories have been explored extensively in recent times. These theories can be classified as degenerate or nondegenerate. If a higher derivative theory is nondegenerate, there exists a ghost-like instability known as Ostrogradsky instability\cite{Ostrogradsky:1850fid} which can be identified with the Hamiltonian of the theory containing unavoidable linear momenta terms\cite{Woodard:2015zca,chen2013higher}. The canonical energy is expressed in phase space variables, where linear momenta terms make the energy unbounded from below. Whereas, in a degenerate theory, the clever cancellation among higher derivative terms provides second-order equations of motion, avoiding Ostrogradsky ghosts. These theories  include  Horndeski theories\cite{Horndeski:1974wa,Nicolis:2008in,Deffayet:2011gz,Kobayashi:2011nu}, beyond Horndeski\cite{Gleyzes:2014dya,Gleyzes:2014qga} and other degenerate higher derivative theories\cite{Langlois:2015cwa,Motohashi:2016ftl,BenAchour:2016fzp} (for a review, refer to\cite{Langlois:2018dxi,Kobayashi:2019hrl}). Mostly these theories include curvature tensor in linear order coupled with higher derivatives of the scalar field. 

Apart from these, theories with higher-order curvature terms are also studied in the context of Chern-Simons gravity\cite{Lue:1998mq, Jackiw:2003pm}, ghost-free Weyl gravity, and ghost-free parity violation theory\cite{Deruelle:2012xv}. Further, the class of theories which are arbitrary nonlinear functions of fully degenerate Lagrangian, L=$f(R, GB, P)$, where R, GB and P represent Ricci scalar, Gauss-Bonnet term and Pontryagin term, respectively, are studied in the context of partially degenerate theories (for more details refer to\cite{Crisostomi:2017ugk}). 

Generally,  the theories in which scalar field is an arbitrary function of space and time coupled non-minimally with gravity are known as general covariant gravity theories.  If they are studied under a well known unitary gauge in which scalar field is only a function of time, comes under spatially covariant theory (SCT) of gravity\cite{Gao:2014soa,Gao:2014fra,Fujita:2015ymn,Gao:2018znj,Gao:2019lpz,Gao:2018izs,Gao:2019liu}. Some examples of SCTs include Horava Lifshitz gravity\cite{Horava:2009uw,Blas:2009qj}, Cuscuton theory\cite{Afshordi:2006ad,Afshordi:2007yx} and its generalization\cite{Gomes:2017tzd,Iyonaga:2018vnu}. The unitary gauge appears naturally in the cosmological background (homogeneous space-time) by the assumption the scalar field has only a time-like gradient. Hence it would be a useful exercise to check whether a particular theory is degenerate under the unitary gauge or not. The class of theories degenerates in the unitary gauge dubbed as U-DHOST, which defines the broader class of degenerate theories than DHOST.  The degeneracy can be checked for the general structure of Lagrangian but the unitary gauge dramatically reduces the effort of checking degeneracy conditions by simplifying the structure of Lagrangian. Recently in \cite{DeFelice:2018ewo, DeFelice:2021hps}, it is shown that in a general coordinate system, the extra mode appearing in U-DHOST theories is non-propagating called shadowy modes.

For the completeness we point out that the recent detection of gravity waves by LIGO observatory puts a constraint on the speed of gravity wave and in turn this constraint can further restrict the parameter space of higher derivative theories ref.\cite{Creminelli:2017sry,Sakstein:2017xjx,Ezquiaga:2017ekz,Baker:2017hug,Amendola:2017orw,Langlois:2017dyl,Ezquiaga:2018btd,Nunes:2018zot,Peirone:2019yjs,Battye:2018ssx, Amendola:2018ltt,Copeland:2018yuh,Cai:2018rzd,Creminelli:2006xe,Cheung:2007st}.  This fact motivates us to include additional higher derivative terms in these theories, which may obtain a larger allowed parameter space from the speed of gravity wave.

In this work, we develop a model by considering the system of Lagrangians containing at most second-order derivatives of the scalar field and their coupling with the linear curvature terms. We separate the second-order time derivative terms in the total Lagrangian using $(3+1)$ decomposition. First, we analyze by taking only the linear curvature terms, but we cannot find any condition for which all the second-order time derivative terms vanish. Therefore, we introduce additional quadratic terms in curvature coupled with a single derivative of the scalar field and a combination of the quartic second derivative of the scalar field.  For a general combination, higher-order derivative terms are still present. Still, in the unitary gauge, we can show that we can get rid of these terms for a particular combination of these terms. 
In other words, we can conclude that the total action is not free from  Ostrogradsky ghosts, but in the unitary gauge, it provides at most a second-order equation of motion for both metric and scalar fields. As an application of our framework, we compute the equation of motion for flat FRW (Friedmann-Lemaitre-Robertson-Walker) metric, which gives rise to an additional correction term in standard FRW equations. 

The organization of this work is as follows: in section 2, we construct the general structure of Lagrangian. In section 3, we review the basics of 3+1 decomposition.  We analyze our Lagrangian in both general and unitary gauge by using 3+1 decomposition in section 4. In section 5, we derive the action for working examples by taking flat, open, and closed FRW metrics. In section 6, we summarize our results.

\section{Construction of Theory}
 Let us start by discussing some important features of Horndeski theories, which  are well described by the following  Lagrangian,
\begin{eqnarray}
L=  L^{H}_{2}+ L^{H}_{3}+ L^{H}_{4}+ L^{H}_{5},
  \end{eqnarray}
where, 
\begin{align}
L^{H}_{2} &=G_{2}(\phi, X),  \\
    L^{H}_{3}& =G_{3}(\phi, X) \Box{\phi},  \\
    L_{4}^{H}& =G_{4}(\phi, X) R - 2G_{4, X}(\phi, X)(   \Box\phi^{2} - \phi_{ab} \phi^{ab}), \\ 
    L_{5}^{H}& =G_{5}(\phi,  X) G_{ab}\phi^{ab}+\frac{1}{3}G_{5}(\phi, X) R - (   \Box\phi^{3} - 3  \Box{\phi} \phi_{ab} \phi^{ab}+ 2 \phi_{ab} \phi_{c}^{b} \phi^{ac}),  
    \end{align}
with following symbols as, $ \phi_{a}= \nabla_{a} \phi $,  $ \phi_{ab}= \nabla_{a} \nabla_{b} \phi $, $ X=\nabla_{a} \phi  \nabla^{a} \phi $.  and $G_{ab}$ is Einstein tensor.\\
The Lagrangian   $L_{2}^{H}$ is the collection of all possible combination scalar fields and its first derivative, and $L_{3}^{H}$ is the direct extension of  $L_{2}^{H}$  by multiplying a second-order derivative term $\Box{\phi} $. However, the combination of  $L_{2}^{H}$ and $L_{3}^{H}$ yields a second-order equation of motion. Further, $L_{4}^{H}$ and $L_{5}^{H}$ contain both first and second derivatives of the metric and scalar field. We also obtain the second-order equation of motion by adding all four terms despite having higher derivative terms, which is possible due to the degeneracy present in this structure.

Now we are interested in considering a particular form of action that contains derivatives of the scalar field at most second order and curvature terms up to quadratic order. For this specific choice, the action can be written as,
\begin{eqnarray}
S&=& \int d^{4}x \sqrt{-g}\ \tilde{A}( \nabla_{\mu} \nabla_{\nu} \phi,\nabla_{\mu}\phi,\phi)+\ \int d^{4}x \sqrt{-g} \tilde{B}(g_{\mu \nu},R_{\mu \nu \rho \sigma},\phi,\nabla_{\mu}\phi)+\ \int d^{4}x \sqrt{-g} \tilde{C}^{\mu \nu ,\rho \sigma}  \nabla_{\mu} \nabla_{\nu} \phi \nabla_{\rho} \nabla_{\sigma} \phi,\label{base} \\\nonumber S&=&S_{1}+S_{2}+S_{3}, 
\end{eqnarray}
where, $\tilde{A}$ is  a function of $\phi$ and its first and second  derivatives,   $\tilde{B}$ takes care of curvature terms upto quadratic order as well as coupling terms between curvature  and  scalar field along with its first order derivatives, and  $\tilde{C}^{\mu \nu \rho \sigma}$ contains linear curvature terms along with $\phi $ and its derivatives. One can get the Horndeski Lagrangian from eq.(\ref{base}) by choosing,  $\tilde{A}=0$, $\tilde{B}$ as just a function of the Ricci scalar, and $\tilde{C}^{\mu \nu\rho \sigma}$ as only a function of $g_{\mu\nu}$, $\phi$ and $\nabla_{\mu}\phi$ (refer to \cite{Langlois:2015cwa} for more details).

In action eq.(\ref{base}), we come across both second derivative in scalar field and metric, which is why we expect the appearance of Ostrogradsky ghosts. 
However, in literature, there are various examples of such higher derivative theories where one can avoid the appearance of Ostrogradsky ghosts by reducing the phase space non-trivially, which leads to the degeneracy condition in these theories \cite{Langlois:2015cwa, Crisostomi:2017ugk, BenAchour:2016fzp, Achour:2016rkg, Langlois:2015skt, Motohashi:2016ftl, Motohashi:2018pxg, Motohashi:2017eya, Langlois:2020xbc}.
To find any existing degeneracy present in our theory, first, we need to separate space and time derivatives by using 3+1 decomposition. In the next section, we briefly review the basics of this technique.
\section{3+1 decomposition}
We have a four dimensional space-time M with metric $ g_{a b} $. This space-time is split into family of non-intersecting space-like 3-surfaces  characterized by metric $ h_{a b}$ (spatial 3-dimensional metric) and these 3-surfaces are connected by defining a normal direction denoted by a vector $ n^{a} $, satisfying the normalization condition $n_{a}n^{a}=-1.$ For more details, please refer to\cite{baumgarte2010numerical}.
In this formalism the 4-dimensional metric can be written in terms of $ h_{a b}$ and normal vector $ n^{a}$,
\begin{equation}
g_{a b}= h_{a b}- n_{a} n_{b}.
\end{equation} 
Any vector $A_{a}$ in (3+1) decomposition can be written as,
\begin{eqnarray}
A_{a}=\mathcal{A}_{a}- A_{*}n_{a},
\end{eqnarray}
where, $\mathcal{A}_{a}$ and  $A_{*}$ are defined as, 
\begin{eqnarray}
\mathcal{A}_{a}=h_{a}^{b} A_{b}, \qquad\qquad A_{*}= A_{a}n^{a}.
\end{eqnarray}
$\mathcal{A}_{a}$ is purely spatial part of $A_{a}$, with the property $n^{a}\mathcal{A}_{a}=0$. 
By using the property $\nabla_{a} A_{b}=\nabla_{b} A_{a}$, the 3+1 decomposition of derivative of $A_{a}$ is given by,
\begin{eqnarray}
\nabla_{a} A_{b} = \mathcal{D}_{a} \mathcal{A}_{b} - A_{*} K_{a b} +n_{a}(K_{b c} \mathcal{A}^{c} -\mathcal{D}_{b} A_{*} )+ n_{b}(K_{a c} \mathcal{A}^{c} -\mathcal{D}_{a} A_{*} )+ \nonumber \\  n_{a} n_{b}(\mathcal{L}_{n}{A}_{*}- \mathcal{A}_{c} a^{c}),\label{main} 
\end{eqnarray}
where $ N^{a}$ is the shift vector, N is the lapse function, $\mathcal{D}_{a}$ denotes spatial derivative and $ K_{a b}$ is extrinsic curvature tensor related to first time derivative of spatial metric,    
\begin{eqnarray} 
K_{a b}=\frac{1}{2N}\left( \dot{h_{a b}}-\mathcal{D}_{a} N_{b}-\mathcal{D}_{b}N_{a}\right), 
\end{eqnarray}
and $\mathcal{L}_{n}{A}_{*}$ which involves second time derivative of scalar field is,
\begin{eqnarray}
 \mathcal{L}_{n}{A}_{*}=\frac{1}{N}(\dot{A_{*}}-N^{c}\mathcal{D}_{c}A_{*}).
\end{eqnarray}

Here we introduce a new set of  variables \{$ U_{a b}$, $ Y_{b}$, $ \dot{Z}_{*}$\} to  further simplify the eq.(\ref{main}), where $ U_{a b}=(\mathcal{D}_{a} \mathcal{A}_{b} - A_{*} K_{a b}), $ behaves as a symmetric spatial tensor, 
$ Y_{b}=(K_{b c} \mathcal{A}^{c} -\mathcal{D}_{b} A_{*}$),  transforms like a spatial vector, and
$ \dot{Z}_{*}=(\mathcal{L}_{n}{A}_{*}- \mathcal{A}_{c} a^{c}),$ contains second time derivative of scalar field. Therefore, in terms of these new variables, eq.(\ref{main}) becomes,
\begin{eqnarray}
(\nabla_{a} A_{b})= \dfrac{1}{N}  n_{a} n_{b} \dot{Z}_{*}-U_{ab}-n_{a}Y_{b} \ -n_{b} Y_{a}.
\label{R1}
\end{eqnarray}
Ricci scalar in (3+1) decomposition takes the form,
\begin{equation}
R= \mathcal{R}+ K^{2}-3 K_{a b} K^{a b}+ 2 h^{a b} \mathcal{L}_{n}  K_{a b} -2 \mathcal{D}_{b} a^{b}-2a_{b} a^{b}, \label{rs}
\end{equation}
 Where $a_{b}$ is acceleration, $\mathcal{R}$ is spatial curvature  and $ \mathcal{L}_{n}K_{a b}$ is Lie derivative of  extrinsic curvature tensor  which contains terms with second order derivative of induced metric $h_{ab},$
\begin{eqnarray}
\mathcal{L}_{n}K_{a b}=\frac{1}{N}(\partial_{t} K_{a b}-N^{c}\mathcal{D}_{c}K_{a b} ).
\end{eqnarray}
Further, we introduce a scalar $\mathcal{F}_{1}$ which takes care of first derivative of spatial metric $h_{ab}$ and purely spatial terms, then eq.(\ref{rs}) can be simplified to,
\begin{equation}
R=  2 h^{a b} \mathcal{L}_{n}  K_{a b} +\mathcal{F}_{1},\label{R2} \,
\end{equation} 
where, 
\begin{equation}
\mathcal{F}_{1}=\mathcal{R}+ K^{2}-3 K_{a b} K^{a b} -2  \mathcal{D}_{b} a^{b}-2a_{b} a^{b}.
\end{equation}
 A similar analysis of  (3+1) decomposition of Ricci Tensor $ R_{a b}$ can be expressed in the following way. 
The purely spatial part of decomposition of Ricci tensor is,
\begin{eqnarray} 
\begin{aligned}
{}_{\perp}R_{ab}&= \mathcal{R}_{a b}+ K_{a b} K - 2K_{a s} K^{s}_{b}+  \mathcal{L}_{n}  K_{a b} -  \mathcal{D}_{(a} a_{b)}- 2a_{a} a_{b}=  \mathcal{L}_{n}  K_{a b}+F_{a b}, \label{ps}
\end{aligned}
\end{eqnarray}
where, \begin{equation}
  F_{a b} =  \mathcal{R}_{a b}+ K_{a b} K - 2K_{a s} K^{s}_{b}-  \mathcal{D}_{(a} a_{b)}- 2a_{a} a_{b},
\end{equation}  which contains only first order derivatives of $h_{ab}$ and purely spatial terms. \\ 
One-normal projection of Ricci tensor denoted by a vector notation $V_{b}$ is,
\begin{eqnarray}
{}_{\perp}R_{b n}= \mathcal{D}_{s}K_{b}^{s}- \mathcal{D}_{b} K = V_{b}. \label{on}
\end{eqnarray}
Finally, two-normal projection of Ricci tensor can be written as,
\begin{eqnarray}
{}_{\perp}R_{n n}=  K_{s t} K^{s t}- h^{s t} \mathcal{L}_{n}  K_{s t} +  \mathcal{D}_{s} a^{s}+a_{s} a^{s}
=\mathcal{F}_{2}-h^{s t} \mathcal{L}_{n}  K_{s t}. \label{tn}
\end{eqnarray}
where,
\begin{equation}
  \mathcal{F}_{2}=   K_{s t} K^{s t} +  \mathcal{D}_{s} a^{s}+a_{s} a^{s}.
\end{equation}
Now the full expression of Ricci tensor by combining eq.(\ref{ps}), eq.(\ref{on}) and eq.(\ref{tn}) becomes,
 \begin{eqnarray}
  R_{a b}=  \mathcal{L}_{n}  K_{a b}+F_{a b}
   -2 n_({a}V_{b)}+n_{a}n_{b}(\mathcal{F}_{2} -h^{s t} \mathcal{L}_{n}  K_{s t}). \label{R3}
\end{eqnarray} 
Similarly all the (3+1) decomposition of Riemann tensor can be written as\cite{baumgarte2010numerical},
\begin{eqnarray}
{}_{\perp} R_{a b c d} = \mathcal{R}_{a b c d}+K_{a c} K_{b d}- K_{a d}K_{b c}. \label{R4}\\
{}_{\perp} R_{a b c n}  = D_{a}K_{b c}-D_{b}K_{a c}, \label{R5}
\end{eqnarray} 
 and 
\begin{eqnarray}
{}_{\perp} R_{a n b n}=  K_{a u} K^{u}_{b}- L_{n}  K_{a b} + D_{(a} a_{b)}+a_{a} a_{b}= \mathcal{F}_{ab}- L_{n}  K_{a b}, \label{R6}
\end{eqnarray}
where,\begin{equation}
   \mathcal{F}_{ab}= K_{a u} K^{u}_{b} + D_{(a} a_{b)}+a_{a} a_{b}.
\end{equation}
The eq.(\ref{R4}), eq.(\ref{R5}) and  eq.(\ref{R6}) are known as Gauss, Codazzi and Ricci relations respectively.
Next we write  various  possible  terms  of eq.(\ref{base}) and their 3+1 deomposition forms.

\subsection{ Construction of $\tilde{C}^{\mu \nu \rho \sigma}$} 
The last term of action  eq.(\ref{base}) is,
 \begin{eqnarray}
 S_{3}= \int d^{4}x \sqrt{-g} \  \tilde{C}^{\mu \nu ,\rho \sigma}  \nabla_{\mu} \nabla_{\nu} \phi \nabla_{\rho} \nabla_{\sigma} \phi \label{S3}.
\end{eqnarray} 
It is already mentioned earlier that $ \tilde{C}^{ \mu \nu \rho \sigma} $ contains only the linear curvature coupling terms, so its possible structure is given by\cite{Joshi:2019tzr},
 \begin{eqnarray}
\tilde{C}^{\mu \nu \rho \sigma}= (D_{1} g^{\mu \rho} g^{\nu \sigma} +D_{2} g^{\mu \sigma} g^{\nu \rho}) R  + (D_{3} g^{ \eta \rho} g^{ \mu \nu } g^{\beta \sigma}  +D_{4}  g^{\mu \eta} g^{\beta \rho} g^{\nu \sigma})R_{\eta \beta} \nonumber \\ + (D_{5}  g^{ \mu \eta } g^{  \nu \beta } g^{\gamma \rho} g^{ \delta \sigma} + D_{6} g^{  \mu \eta} g^{ \sigma \beta } g^{\gamma \rho} g^{\delta \nu}) R_{\eta \beta \gamma \delta}, 
\end{eqnarray}
where $D_{i}(\phi,X) s(i=1,2,..6)$ are general function of $\phi $ and  its first derivative.
Using Lagrange multiplier $\lambda_{\mu}$ and replacing $\nabla_{\mu} \phi$ to a new $A_{ \mu}$ field, the  eq.(\ref{S3}) can be rewritten as,
\begin{eqnarray}
\begin{aligned}
 S_{3}&= \int d^{4}x \sqrt{-g} \  \tilde{C}^{\mu \nu ,\rho \sigma} \nabla_{\mu} A_{\nu}  \nabla_{\rho} A_{\sigma}+\lambda^{\mu} (\nabla_{\mu} \phi-A_{\mu}),\\&=\int d^{4}x \sqrt{-g} \ (L_{1}+L_{2}+L_{3}+L_{4}+L_{5}+L_{6})+\lambda^{\mu} (\nabla_{\mu} \phi-A_{\mu}),\label{simple}
 \end{aligned}
\end{eqnarray}
where $L_i$s  (i=1,2...6) are defined as,
\begin{eqnarray}
\begin{aligned}
L_{1}&=D_{1} R  g^{c e} g^{d f}  \nabla_{c} A_{d} \nabla_{e} A_{f},\label{l16} &\\
L_{2}&= D_{2}R g^{c d} g^{e f}  \nabla_{c} A_{d}  \nabla_{e} A_{f},&\\
L_{3}&=D_{3} g^{a c} g^{b e} g^{d f} R_{a b}  \nabla_{c} A_{d} \nabla_{e} A_{f}, \\
L_{4}&=D_{4} g^{a e} g^{c d} g^{b f} R_{a b} \nabla_{c} A_{d}   \nabla_{e} A_{f},&\\
L_{5}&= D_{5}g^{c a} g^{d b} g^{l e} g^{m f} R_{a b l m} \nabla_{c} A_{d} \nabla_{e} A_{f}, &\\
L_{6}&= D_{6}g^{c a} g^{f b} g^{l e} g^{m d} R_{a b l m}\nabla_{c} A_{d} \nabla_{e} A_{f}.&
\end{aligned}
\end{eqnarray}

\subsection{Representation of action $S_3$ in 3+1 decomposition} 
We decompose Lagrangians $L_{1}$ to $L_{6}$, by using 3+1 decomposed  relation of  $ \nabla_{c} A_{d}$ in eq.(\ref{R1}), and the curvature tensor provided in eq.(\ref{R2}), (\ref{R3})-(\ref{R6}). Then we separate out second derivative  of metric with corresponding coefficient in each Lagrangian, which yields,
\begin{eqnarray}
 L_{1-4}=D_{1}\left[ \left(2 h^{a b} \mathcal{L}_{n}  K_{a b}+ \mathcal{F}_{1} \right)\left(\dot{Z}_{*}^2- 2 Y_{c}Y^{c}+  U_{cd} U^{cd}\right)\right] \nonumber\\
 \label{S1a}
+D_{2}\left[\left(2 h^{a b} \mathcal{L}_{n}  K_{a b}+ \mathcal{F}_{1} \right)\left(\dot{Z}_{*}^2 -2  \dot{Z}_{*}  U+U^{2}\right) \right] \nonumber\\
  +D_{3}\Bigg[\left(-h^{a b}\mathcal{L}_{n}K_{a b} +\mathcal{F}_{2} \right)  \left(\dot{Z}_{*}^2+Y_{d}Y^{d}\right)+\left(\mathcal{L}_{n}K_{a b}+F_{a b}\right)\\ \nonumber\left( U_{d}^{b} U^{ad}- Y^{a}Y^{b}\right)  + 2 V_b \left(\dot{Z}_{*} Y^{b}-U_{a}^{b}Y^{a}\right)\Bigg] \nonumber\\
+D_{4}\Bigg[ \left(-h^{a b} \mathcal{L}_{n}  K_{a b}+\mathcal{F}_{2}\right)\left(\dot{Z}_{*}^2+\dot{Z}_{*}U \right)+\left( \mathcal{L}_{n}  K_{a b}+F_{a b}\right)\nonumber\\ \nonumber\left( U U^{a b}-\dot{Z_{*}} U^{a b}\right) +2V_b \left(\dot{Z}_{*}Y^{b}-UY^{b}\right)\Bigg],
\end{eqnarray}
where $L_{1-4}$ is  the linear sum of  Lagrangian's $L_{1}$ to $L_{4}$ and throughout the paper we will use this type of  notation for  denoting linear sum. $L_{5}$ does not contribute 
due to antisymmetric properties of Riemann tensor. The 3+1 decomposed form of  Lagrangian $L_{6}$ is given as,
\begin{eqnarray}
L_{6}=D_{6}\bigg[2\mathcal{L}_{n}K_{ab}\left(\dot{Z}_{*}U^{a b}-Y^{a}Y^{b}\right) +2\mathcal{F}_{ab}(Y^{a}Y^{b}-\dot{Z}_{*}U^{a b})-{}_{\perp} R_{a b c d}U^{ac}U^{bd}\nonumber\\+4U^{bc}Y^{a}\left(\mathcal{D}_{c}K_{ab}-\mathcal{D}_{a}K_{bc}\right)\bigg].\label{s2}
\end{eqnarray}
These families of Lagrangian are equipped with coupling between the first and second-order derivatives of both scalar field and metric. The most problematic part is the coupling of the quadratic time derivative of the scalar field ($\dot{Z}_{*}^2$) to others. To tackle this problem, we separate these types of  terms with their corresponding coefficient, then the eq.(\ref{S1a}) and eq.(\ref{s2}) take the form,
\begin{eqnarray} 
L_{1-6} =  \left(2h^{a b} \mathcal{L}_{n}   K_{a b}+ \mathcal{F}_{1} \right) \dot{Z}_{*}^2\left(D_{1}+D_{2}\right)+
\left(h^{a b}\mathcal{L}_{n} K_{a b} +\mathcal{F}_{2}\right) \dot{Z}_{*}^2\left(D_{3}+D_{4}\right)+
h^{a b}\mathcal{L}_{n} K_{a b}Y_{c}Y^{c}
\nonumber\\
\left(-4D_1-D_3\right)+h^{ab}\mathcal{L}_{n} K_{a b} \dot{Z}_{*} U\left( -4D_{2}-D_4\right)+\mathcal{L}_{n} K_{a b}Y^{a}Y^{b}\left(-D_{3}-2D_{6}\right)
\nonumber\\
+\left(2h^{a b} \mathcal{L}_{n} K_{a b}+\mathcal{F}_{1}\right)\left( D_{1} U^{cd} U_{cd}+D_{2}U^2\right) +\mathcal{L}_{n} K_{a b}\dot{Z}_{*}  U^{ab}\left( -D_4+2D_6\right) 
\nonumber\\
+ \left(\mathcal{L}_{n} K_{a b}+F_{ab}\right)\left(D_{3} U_{d}^{a} U^{ b d} +D_{4}U U^{a b}\right)+Y^{d}Y_{d}\left(-2D_{1}\mathcal{F}_{1}+D_{3} \mathcal{F}_{2}\right)
\nonumber\\
+\dot{Z}_{*}U\left(-2D_{2}\mathcal{F}_{1}+D_{4}\mathcal{F}_{2} \right) + 2 V_b \left[ \dot{Z}_{*} Y^{b}\left(D_{3}+D_{4}\right)-D_{3}U_{a}^{b}Y^{a}-D_{4}UY^{b}\right]
\nonumber\\
+Y^{a}Y^{b}(-F_{ab}D_{3}+2D_{6}\mathcal{F}_{ab}+\dot{Z}_{*}U^{a b}(-F_{ab}D_{4}-2D_{6}\mathcal{F}_{ab})
\nonumber\\
-{}_{\perp} R_{a b c d}U^{ac}U^{bd}+4U^{bc}Y^{a}\left(\mathcal{D}_{c}K_{ab}-\mathcal{D}_{a}K_{bc}\right).
 \end{eqnarray}
 
It is easy to see that we can tune the coefficients $D_1-D_6$ in such a way that coupling of  $\dot{Z}_{*}^2$ to other terms is removed. Consequently, we obtain the following conditions,
\begin{align}
D_1+D_2=0, \nonumber\\D_3+D_4=0,  \nonumber\\ -4D_1-D_3=0,  \nonumber\\ -4D_2-D_4=0,  \nonumber\\ -D_3-2D_6=0, \nonumber\\ -D_4+2D_6=0.\label{con1}
\end{align}
After solving these equations, we get \begin{eqnarray}
   D_1=-D_{2}= -\dfrac{D_3}{4}=\dfrac{D_4}{4}=\dfrac{D_{6}}{2}.\label{sigma} \end{eqnarray} 
After applying these conditions, $L_{1-6}$ becomes,
 \begin{eqnarray}
L_{1-6}= D_{1}\bigg[\left(U^{cd}U_{cd}-U^{2}\right)\left(2h^{ab}\mathcal{L}_{n}K_{ab}+\mathcal{F}_{1}\right)+4\left(U U^{ab}-U^{a}_{d}U^{db}\right)\left(\mathcal{L}_{n}K_{ab}+F_{ab}\right)
\\ \nonumber+
2\dot{Z}_{*}U\left(\mathcal{F}_{1}+2\mathcal{F}_{2} \right)-4\dot{Z}_{*}U^{a b}(F_{ab}+\mathcal{F}_{ab})-2
Y^{d}Y_{d}\left(\mathcal{F}_{1}+2 \mathcal{F}_{2}\right)++
 8 V_b \left(U_{a}^{b}Y^{a}-UY^{b}\right)\\ \nonumber+4
Y^{a}Y^{b}(F_{ab}+\mathcal{F}_{ab})
-2{}_{\perp} R_{a b c d}U^{ac}U^{bd}+8U^{bc}Y^{a}\left(\mathcal{D}_{c}K_{ab}-\mathcal{D}_{a}K_{bc}\right)\Bigg\} \Bigg]. \label{rip} 
 \end{eqnarray}
 After removing $\dot{Z}_*^2$ terms we are still left with the linear terms in second-order derivative of both scalar field and metric. After putting the values of $ U_{ab}$, $Y_{b}$ and  $ \dot{Z}_{*}$, we obtain,
\begin{eqnarray}
\begin{aligned}
L_{1-6}= & D_{1}\bigg[(2h^{ab} \mathcal{L}_n K_{ab}+\mathcal{F}_{1}) \bigg\{ A_{*}^2 K_{cd} K^{cd} -A_{*}^2 K^{2} -  A_{*} \ K^{cd} (D_{c}\mathcal{A}_{d}) +  A_{*} K (D^{c}\mathcal{A}_{c})- A_{*} K_{cd} (D^{c}\mathcal{A}^{d}) 
 \\&+  (D_{c}\mathcal{A}_{d})(D^{c}\mathcal{A}^{d}) +  A_{*} K (D^{d}\mathcal{A}_{d}) -(D^{c}\mathcal{A}_{c}) (D^{d}\mathcal{A}_{d})\bigg\} -4(\mathcal{L}_n K_{bc}+F_{bc})\bigg\{A_{*}^2 K_{a}{}^{c} K^{ab} -A_{*}^2 K K^{bc}
  \\&
  -  A_{*} K^{ab} (D_{a}\mathcal{A}^{c}) +A_{*} K^{bc} (D^{a}\mathcal{A}_{a}) -   A_{*} K_{a}{}^{c} (D^{a}\mathcal{A}^{b}) +   (D_{a}\mathcal{A}^{c})(D^{a}\mathcal{A}^{b}) + A_{*} K (D^{b}\mathcal{A}^{c)}
 \\&
-(D^{a}\mathcal{A}_{a})( D^{b}\mathcal{A}^{c})\bigg\}+ 2\dot{Z}_{*}\bigg\{2( F^{ab} + \mathcal{F}^{ab}) (A_{*} K_{ab} -  (D_{a}\mathcal{A}_{b})) +( \mathcal{F}_{1} + 2 \mathcal{F}_{2}) (- \ A_{*} K + (D^{a}\mathcal{A}_{a})\bigg\}
\\&
-2\bigg\{(\mathcal{F}_{1}+2 \mathcal{F}_{2})(K_{ca}\mathcal{A}^{c}-(\mathcal{D}_{a}A_{*}))(K_{d}^{a}\mathcal{A}^{d}-(\mathcal{D}^{a}A_{*})) + 2( F^{ab} -  \mathcal{F}^{ab})(K_{ca}\mathcal{A}^{c}-(\mathcal{D}_{a}A_{*}))(K_{db}\mathcal{A}^{d}
\\&
-\left(\mathcal{D}_{b}A_{*}\right) ) + 4 V^{a} (K_{d}^{b}\mathcal{A}^{d}-(\mathcal{D}^{b}A_{*}))(- A_{*} K_ {ab} + (D_ {a} \mathcal{A}_ {b}) ) -4{}_{\perp} R_{a \ \ n}^{\ bc}(K_{d}^{a}\mathcal{A}^{d}-(\mathcal{D}^{a}A_{*})) 
(- A_{*} K_ {bc} + ( D_ {b} \mathcal{A}_ {c}))
\\&- 
 4 V^{a}(K_{ca}\mathcal{A}^{c}-(\mathcal{D}_{a}A_{*})) (- A_{*} K +( D^{b} \mathcal{A}_ {b})) -  {\perp} R_{acbd} (- A_{*} K^{ab} + (D^{a} \mathcal{A}^{b})) (- A_{*} K^{cd} + (D^{c} \mathcal{A}^{d}))\bigg\}\Bigg].\label{huged}
 \end{aligned} 
 \end{eqnarray} 
This expression includes $ K_{ab}$, $A_{*}$, $\mathcal{A}^{a} $  and their spatial derivatives with linear second-order derivatives of both metric and scalar fields.  We can not find a particular choice of $D_1-D_{6}$ (unless all $D_1-D_{6}$ are zero), which can cancel the contribution of the linear second derivative of the metric as well as that of the scalar field.

\subsection{Construction of $\tilde{B}$} 
To remove linear higher derivative terms of metric in eq.(\ref{huged}),
 here we introduce a structure for $\tilde{B}$ by allowing all possible terms of quadratic  curvature coupled with the first derivatives of the scalar field, which are
\begin{align}
L_{7} &= D_{7}g^{a b} g^{c d} g^{e f}  R_{a b}R_{c d}A_{e} A_{f}, \nonumber\\
L_{8} &= D_{8}g^{a b} g^{c e} g^{d f}  R_{a b}R_{c d}A_{e} A_{f},
\nonumber\\ L_{9} & = D_{9}g^{a c} g^{b d} g^{e f}  R_{a b}R_{c d}A_{e} A_{f}, 
\nonumber\\L_{10} & = D_{10}g^{a c} g^{b e} g^{d f}  R_{a b}R_{c d}A_{e} A_{f},
\nonumber\\L_{11} &= D_{11}g^{c r} g^{q b} g^{d f} g^{s e} g^{a p} R_{a c b d}R_{p q r s}A_{e} A_{f}.\label{eqsol}
\end{align}
 Recall that the higher derivative of metric $L_{n}K_{ab}$ terms in the expression $L_{1-6}$ eq.(\ref{huged}) appear  with  a combination $ (K_{ab}$, $A_{*}$, ${A}^{a})$ and spatial derivatives. We would expect terms with such a combination to emerge from $L_7-L_{11}$ by which all terms with quadratic derivatives of the metric vanish.
 By  performing 
 3+1 decomposition of Lagrangians $L_{7-11}$, we obtain,
\begin{eqnarray}
\begin{aligned}
  L_{7-11}=& D_{7}\bigg[\Big\{4 h^{a b} h^{c d}\left(\mathcal{L}_{n} K_{a b}\right) \left(\mathcal{L}_{n}  K_{c d}\right)+
  4\mathcal{F}_{1} h^{a b} \left(\mathcal{L}_{n}  K_{a b}\right)
  + \mathcal{F}_{1}\mathcal{F}_{1}\Big\}\left(\mathcal{A}^{2}-A_{*}^{2}\right)\bigg]
  \\&+
  D_{8}\bigg[-2A_{*}^{2}h^{ab}\mathcal{L}_{n}K_{ab}h^{cd} \mathcal{L}_{n}K_{cd}+
  2\mathcal{A}^{a}\mathcal{A}^{b}h^{cd}\mathcal{L}_{n}K_{ab}\mathcal{L}_{n}K_{cd} +
  A_{*}^{2}h^{ab}\mathcal{L}_{n}K_{ab}\left(-\mathcal{F}_{1}+
  2\mathcal{F}_{2}\right) 
\\&
+h^{cd}\mathcal{L}_{n}K_{cd}\left(\mathcal{A}^{a}\mathcal{A}^{b}F_{ab}-
4A_{*}\mathcal{A}^{a}V_{a}\right)+
  \mathcal{F}_{1}\left(\mathcal{A}^{c}\mathcal{A}^{d}\mathcal{L}_{n}K_{cd}-
  2A_{*}\mathcal{A}^{c}V_{c}\right)+\mathcal{F}_{1}\mathcal{A}^{a}\mathcal{A}^{b}F_{ab}+A_{*}^{2}
  \mathcal{F}_{1}\mathcal{F}_{2}\bigg]\\&
   +D_{9}\bigg[\bigg\{h^{ab}h^{cd}\mathcal{L}_{n}K_{ab} \mathcal{L}_{n}K_{cd}+
   h^{ab}h^{cd}\mathcal{L}_{n}K_{ac}\mathcal{L}_{n}K_{bd}+2F^{ab}\mathcal{L}_{n}K_{ab}-
   2\mathcal{F}_{2} h^{ab}\mathcal{L}_{n}K_{ab}\\&+ F_{ab}F^{ab}-2h^{ab}V_{a}V_{b}+\mathcal{F}_{2}\mathcal{F}_{2}\bigg\}\left(\mathcal{A}^{2}-A_{*}^{2}\right)\bigg]
  \\&+
    D_{10}\bigg[-A_{*}^{2}h^{ab}\mathcal{L}_{n}K_{ab}h^{cd} \mathcal{L}_{n}K_{cd}+
\mathcal{A}^{a}\mathcal{A}^{b}h^{cd}\mathcal{L}_{n}K_{ac}\mathcal{L}_{n}K_{bd}+
2\mathcal{A}^{a}\mathcal{A}^{b}h^{cd}F_{ad}\mathcal{L}_{n}K_{bc}
\\&
-2A_{*}\mathcal{A}^{a}h^{cd}V_{a}\mathcal{L}_{n}K_{cd}-2A^{a}A_{*}V^{b}\mathcal{L}_{n}K_{ab}+2A_{*}^{2} \mathcal{F}_{2}h^{cd}\mathcal{L}_{n}K_{cd}-(\mathcal{A}^{a}\mathcal{A}^{b}-h^{ab})A_{*}^{2}V_{a}V_{b}
\\&
-2A_{*}\mathcal{A}^{a}h^{bc}V_{c}F_{ab}+\mathcal{A}^{a}\mathcal{A}^{b}h^{dc}F_{ad}F_{bc}+2\mathcal{F}_{2}A_{*}\mathcal{A}^{a}V_{a}-A_{*}^{2}\mathcal{F}_{2}\mathcal{F}_{2} \bigg] \\&
  +D_{11}\bigg[-A_{*}^{2}h^{ab}\mathcal{L}_{n}K_{ab}h^{cd} \mathcal{L}_{n}K_{cd}+
\mathcal{A}^{a}\mathcal{A}^{b}h^{cd}\mathcal{L}_{n}K_{ac}\mathcal{L}_{n}K_{bd}-
2\mathcal{A}^{a}\mathcal{A}^{b}h^{cd}F_{ad}\mathcal{L}_{n}K_{bc}
\\&
+4A_{*}\mathcal{A}^{a} {}_{\perp}\mathcal{R}^{bc}_{\ \ an}h_{c}^{d}
\mathcal{L}_{n}K_{bd}+
2A_{*}^{2} \mathcal{F}^{ab}\mathcal{L}_{n}K_{ab}+\mathcal{A}_{a}\mathcal{A}^{b} {}_{\perp}\mathcal{R}^{acdp} {}_{\perp}\mathcal{R}_{bdcp}
+2A_{*}\mathcal{A}^{a} {}_{\perp}\mathcal{R}^{bcdn} {}_{\perp}\mathcal{R}_{acbd}
\\&
+2A_{*}\mathcal{A}^{b} {}_{\perp}\mathcal{R}_{an}^{\ \ cd} {}_{\perp}\mathcal{R}_{cdbn}+\mathcal{A}^{a}\mathcal{A}^{b}\mathcal{F}_{a}^{c}F_{bc}
+A_{*}^{2}\mathcal{R}_{ancb}\mathcal{R}^{abcn}
-4A_{*}\mathcal{A}^{a} {}_{\perp}\mathcal{R}^{bc}_{\ \ an}\mathcal{F}_{bc}-A_{*}^{2}\mathcal{F}_{ab}\mathcal{F}^{ab} \bigg]. \label{lea}
\end{aligned}
\end{eqnarray}
 $L_{7-11}$ contains terms quadratic and linear in second derivative of metric.
 However, for this set of Lagrangians, we are unable to find any nontrivial relation among different coefficients $D_{7}-D_{11}$ so that terms having quadratic second derivative of metric vanish.

Next, we set a structure of quartic second-order derivative of $\phi$ with the hope that it may help us to remove all linear second derivatives of the scalar field of the Lagrangian $L_{1-6}$.

\subsection{Construction of $\tilde{A}$}
 
In order to obtain terms similar to  linear second-order derivative of scalar field, as in eq.(\ref{huged}), we propose a structure containing four double derivative terms which is,

\begin{eqnarray}
\begin{aligned}
L_{12-16}=& L_{12}+L_{13}+L_{14}+L_{15}+L_{16}, \\
L_{12-16}=&\bigg[D_{12}g^{ab} g^{cd} g^{ef} g^{pq}+D_{13} g^{ae} g^{dp} g^{cf} g^{qb}+D_{14}g^{ac} g^{bd} g^{ep} g^{qf}+D_{15}g^{ab} g^{cd} g^{pf} g^{qe}\\&+D_{16}g^{ab} g^{cq} g^{pf} g^{de} \bigg] \nabla_{a} A_{b}\nabla_{c} A_{d} \nabla_{e} A_{f} \nabla_{p} A_{q} \label{sf}.
\end{aligned}
\end{eqnarray}

It is obvious that here we will encounter terms with higher powers of $\dot{Z}_*$($\dot{Z}^4,$ $\dot{Z}^3,$ and $\dot{Z}^2$) apart form linear terms in $\dot{Z_*}$.

Using 3+1 decomposition $L_{12-16}$ becomes,

\begin{eqnarray}
\begin{aligned}
L_{12-16}&=\dot{Z}_{*}^4\left(D_{12} + D_{13}  + D_{14} + D_{15} + D_{16}\right) + 4 \dot{Z}_{*}^3 U\left(-4 D_{12}  - 2 D_{15} - D_{16}\right)
\\& +  \dot{Z}_{*}^2 U^{2}\left(6 D_{12}+ D_{15}\right)- 4 \dot{Z}_{*}^2 Y_{a} \ Y^{a}\left(-4 D_{13}  + 4  D_{14} - 2 D_{15}- 3 D_{16}\right)
\\& +2 \dot{Z}_{*}^2 U_{bc} U^{bc}\left(2 D_{14}+ D_{15}\right)+D_{12}\bigg[- 4 \dot{Z}_{*} U^{3} + U^{4}\bigg]+D_{13}\bigg[ U_{a}{}^{c} U^{ab} U_{b}{}^{d} U_{cd} 
\\&+ 4 \dot{Z}_{*} U_{ab} Y^{a} Y^{b} - 4 U_{a}{}^{c} U_{bc} Y^{a} \ Y^{b} + 2 Y_{a} Y^{a} Y_{b} Y^{b} \bigg]+D_{14}\bigg[ U_{bc} U^{bc} U_{fp} U^{fp}  
\\& - 4 U_{cf} U^{cf} Y_{b} Y^{b} + 4 Y_{b} Y^{b}Y_{c} Y^{c}\big] +D_{15}\big[- 2 \dot{Z}_{*} U U_{de} U^{de} +  U^2 \ U_{ef} U^{ef}  
\\&+ 4 \dot{Z}_{*} U \ Y_{b} Y^{b} - 2 U^2  Y_{b} Y^{b}\bigg]+D_{16}\bigg[ -  \dot{Z}_{*} U_{b}{}^{d} U^{bc} U_{cd} + \
U U_{c}{}^{e} U^{cd} U_{de}  \\& + 3 \dot{Z}_{*} \
U^{c}{}_{c} Y_{b} Y^{b} + 3 \dot{Z}_{*} U_{bc} Y^{b} Y^{c} - 3 U_{bc} \
U^{d}{}_{d} Y^{b} Y^{c}\bigg].
\end{aligned}
\end{eqnarray} 
From the above equation, it is evident that, the   $\dot{Z_*}^4$ and $\dot{Z_*}^3$ terms vanish if  following conditions on the coefficients from $D_{12}$ to $D_{16} $ are satisfied. Conditions are,
\begin{eqnarray}
D_{12} + D_{13}  + D_{14} + D_{15} + D_{16} =0,\nonumber\\
-4 D_{12}  - 2 D_{15} - D_{16}=0,\nonumber\\
2 D_{14}+ D_{15}=0,\nonumber\\
6 D_{12}  + D_{15}=0,\nonumber\\
-4 D_{13}  - 4  D_{14} - 2 D_{15}- 3 D_{16}=0.\label{con3}
\end{eqnarray}
From eq.(\ref{con3}), we get,
\begin{equation}
     D_{15} =-6D_{12}= D_{13}=-2 D_{14}=-\frac{3}{4} D_{16}.
\end{equation}
After imposing the conditions in eq.(\ref{con3}), $L_{12-16}$ becomes linear in $\dot{Z}_{*}$ and takes the form,
\begin{eqnarray}
 L_{12-16}=D_{15} \bigg[{4 \dot{Z}_{*} U_{a}{}^{c} U^{ab} U_{bc} -
2\dot{Z}_{*} U U_{bc} U^{bc}} + U_{a}{}^{c} U^{ab} \
U_{b}{}^{d} U_{cd} -  \tfrac{4}{3} U U_{b}{}^{d} U^{bc} \
U_{cd}\nonumber\\ + 2 \dot{Z}_{*} U^{3} 
-  \tfrac{1}{2} U_{ab} U^{ab} U_{cd} U^{cd} + U^{2} \
U_{cd} U^{cd} -  \tfrac{1}{6} U^{4}+ 2 U_{bc} U^{bc} Y_{a} Y^{a}\nonumber \\ - 2 U^{2} \
Y_{a} Y^{a} - 4 U_{a}{}^{c} U_{bc} Y^{a} Y^{b} + 4 U_{ab} U \
Y^{a} Y^{b}\bigg].\label{SE}
\end{eqnarray}

Our main aim of introducing such quartic terms was to remove all $\dot{Z}_{*}$ and $\mathcal{L}_{n}K_{ab}$ terms from  $L_{1-6}$, but for the whole Lagrangian  $L_{1-16}$, we can observe from eq.'s (\ref{huged}),(\ref{lea}) and (\ref{SE}) that it is difficult to find a non trivial condition in $D_{i}$'s so that all  $\dot{Z}_{*}$ and $\mathcal{L}_{n}K_{ab}$ terms vanish. Finally, we are left with linear higher derivative terms in scalar field and metric. Therefore, next, we shall analyze our action in a unitary gauge.  

\subsection{Unitary Gauge} 
Here, we look for hidden degeneracy (if any) present in the Lagrangian $L_{1-16}$ in unitary gauge. Unitary gauge is simply characterized by the condition,
 \begin{equation}
     \phi(x,t)=\phi_{o}(t).
 \end{equation} 
 For  more details about unitary gauge refer to     \cite{Gleyzes:2014dya,Gleyzes:2014qga,Langlois:2015cwa}. It is clear that $\mathcal{A}^{a} $ vanishes in this gauge. Then, the structure of variables that appeared in eq.(\ref{R1}) becomes,
 \begin{eqnarray}
   U_{a b}= - A_{*} K_{a b},\ \ \ Y_{b}=0,\ \ \dot{Z}_{*}=\mathcal{L}_{n}A_{*},
 \end{eqnarray} 
and by virtue of this, eq.(\ref{rip}) takes the following form,
 \begin{eqnarray}
 \begin{aligned}
L_{1-6}=& D_{1}\bigg[A_{*}^{2}\left(K^{cd}K_{cd}-K^{2}\right)\left(2h^{ab}\mathcal{L}_{n}K_{ab}+\mathcal{F}_{1}\right)+4A_{*}^{2}\left(K K^{ab}-K^{a}_{d}K^{db}\right)\left(L_{n}K_{ab}+F_{ab}\right)
\\& -2\left(\mathcal{R}+K^{2}-K_{ab}K^{ab}\right) A_{*}K(\mathcal{L}_{n}A_{*})+4A_{*}(\mathcal{L}_{n}A_{*})K^{a b}(\mathcal{R}_{ab}+KK_{ab}-K_{ac}K_{b}^{c})
\\&-2A_{*}^{2}{}_{\perp} R_{a b c d}K^{ac}K^{bd} \Bigg].  \label{smilar}
 \end{aligned}
 \end{eqnarray}
 As we can see that $L_{1-6}, $ contains only linear double derivatives of the scalar field and metric. Similarly in this gauge $L_{7-11}$ eq.(\ref{lea}) takes the following form,
\begin{eqnarray}
  L_{7-11} = D_{7}\bigg[-4A_{*}^{2} h^{a b} h^{c d}\mathcal{L}_{n} K_{a b} \mathcal{L}_{n}  K_{c d}-4A_{*}^{2}
  \mathcal{F}_{1} h^{a b} L_{n}  K_{a b}
   -A_{*}^{2}\mathcal{F}_{1}\mathcal{F}_{1}\bigg]\qquad\qquad\qquad\qquad\nonumber\\+
  D_{8}\bigg[-2A_{*}^{2}h^{ab}\mathcal{L}_{n}K_{ab}h^{cd} \mathcal{L}_{n}K_{cd} +
  A_{*}^{2}h^{ab}\mathcal{L}_{n}K_{ab}\left(-\mathcal{F}_{1}+
  2\mathcal{F}_{2}\right)+A_{*}^{2}
  \mathcal{F}_{1}\mathcal{F}_{2}\bigg]\nonumber\\
   +D_{9}\bigg[-A_{*}^{2}h^{ab}\mathcal{L}_{n}K_{ab}h^{cd} \mathcal{L}_{n}K_{cd}-A_{*}^{2}
   h^{ab}h^{cd}\mathcal{L}_{n}K_{ac}\mathcal{L}_{n}K_{bd}-2A_{*}^{2}F^{ab}\mathcal{L}_{n}K_{ab}\qquad\qquad\qquad\nonumber\\+
   2A_{*}^{2}\mathcal{F}_{2} h^{ab}\mathcal{L}_{n}K_{ab}-A_{*}^{2} F^{ab}F_{ab}+2A_{*}^{2}h^{ab}V_{a}V_{b}-A_{*}^{2}\mathcal{F}_{2}\mathcal{F}_{2}\bigg]\qquad\quad\qquad
  \nonumber\\+
    D_{10}\bigg[-A_{*}^{2}h^{ab}\mathcal{L}_{n}K_{ab}h^{cd} \mathcal{L}_{n}K_{cd}+2A_{*}^{2} \mathcal{F}_{2}h^{ab}\mathcal{L}_{n}K_{ab}+ 2A_{*}^{2}h^{ca}V_{a}V_{c}-A_{*}^{2}\mathcal{F}_{2}\mathcal{F}_{2} \bigg]\nonumber\\
 +D_{11}\bigg[-A_{*}^{2}
   h^{ab}h^{cd}\mathcal{L}_{n}K_{ac}\mathcal{L}_{n}K_{bd}+2A_{*}^{2}h^{bd}h^{ca}\mathcal{F}_{bd}\mathcal{L}_{n}K_{ab}-A_{*}^{2}\mathcal{F}_{ab}\mathcal{F}^{ab}\bigg].\qquad\quad\qquad
  \nonumber\\ \label{fid}
\end{eqnarray}

It can be noticed from eq.(\ref{fid}) that we can separate quadratic second order derivatives of metric through a particular combination of coefficients $D_{7}-D_{11}$ in $L_{7-11}$, unlike the previous analysis in the absence of unitary gauge. After some rearrangement, eq.(\ref{fid}) yields,
\begin{eqnarray}
L_{7-11}= A_{*}^2\bigg[-h^{ab} h^{cd} \
\mathcal{L}_n K_{ac}\mathcal{L}_n K_{bd}( D_{11} + D_{9})  +   h^{ab} h^{cd}\mathcal{L}_n K_{ab}\mathcal{L}_n K_{cd}( 4 D_{7} +2 D_{8} + \
D_{9}+D_{10})\nonumber\\+ h^{a b} \mathcal{L}_{n}  K_{a b}\bigg\{-\mathcal{F}_{1}\left(4D_{7}+D_{8}\right)+2\mathcal{F}_{2}(D_{8}+D_{9}+D_{10})\bigg\}
+2h^{bd}h^{ca}\mathcal{L}_{n}K_{bc}(-D_{9}F_{bd}+
\nonumber\\
\mathcal{F}_{bd}D_{11})- D_{7}\mathcal{F}_{1}\mathcal{F}_{1}  + D_{9}  \mathcal{F}^{ab}\mathcal{F}_{ba} + D_{8}\mathcal{F}_{1}\mathcal{F}_{2} -  D_{10}  \
\mathcal{F}_{2}\mathcal{F}_{2} -D_{9} \mathcal{F}_{2}\mathcal{F}_{2} -  D_{9} F_{a}{}^{c} F^{ab} h_{bc} + \nonumber\\D_{10} \
V_{a} V^{a} + 2 D_{9}  V_{a} V^{a} -  D_{9}   {\perp} R_{acbn} \ {\perp} R^{abcn}\bigg].\label{tu}
\end{eqnarray}
Next, we obtain the conditions under which all quadratic second order derivative term of metric disappear:

\begin{align}
4D_{7}+2D_{8}+D_{9}+D_{10}=0, \nonumber \\ D_{9}+D_{11}=0. \label{con2}
\end{align} 
 By using eq.(\ref{con2}),  the eq.(\ref{tu}) reduces to, 
\begin{eqnarray}
L_{7-11}=A_{*}^{2}\bigg[ -\left(4D_{7}+D_{8}\right) A_{*}^{2} h^{a b} \mathcal{L}_{n}  K_{a b}\bigg\{\mathcal{F}_{1}+2\mathcal{F}_{2}\bigg\}
-2D_{9}h^{bd}h^{ca}\mathcal{L}_{n}K_{bc}(F_{bd}+
\mathcal{F}_{bd})- \nonumber\\D_{7}\mathcal{F}_{1}\mathcal{F}_{1}  + D_{9}  \mathcal{F}^{ab}\mathcal{F}_{ba} + D_{8}\mathcal{F}_{1}\mathcal{F}_{2} -  D_{10}  \
\mathcal{F}_{2}\mathcal{F}_{2} -D_{9} \mathcal{F}_{2}\mathcal{F}_{2} -  D_{9} F_{a}{}^{c} F^{ab} h_{bc} + \nonumber\\D_{10} \
V_{a} V^{a} + 2 D_{9}  V_{a} V^{a} -  D_{9}   {\perp} R_{acbn} \ {\perp} R^{abcn}\bigg]. \label{mah}
\end{eqnarray}
 After putting values of $\mathcal{F}_{1}, \mathcal{F}_{2},F_{bd}$ and $\mathcal{F}_{bd}$ as defined in \textsection III, eq.(\ref{mah})  becomes,
\begin{eqnarray}
\begin{aligned}
L_{7-11}&= \left(4D_{7}+D_{8}\right) A_{*}^{2} h^{a b} \mathcal{L}_{n}  K_{a b}\left(\mathcal{R}+K^{2}-K_{ab}K^{ab}\right)\\&-2D_{9}A_{*}^{2}h^{bd}h^{ca}\mathcal{L}_{n}K_{cd}(\mathcal{R}_{ab}
+ KK_{ab}-K_{ac}K_{b}^{c})+L(I),\label{su}
\end{aligned}
\end{eqnarray}
where we introduce notation L(I) which takes care of all the terms containing spatial derivatives and first time derivative of metric.
Combining equations (\ref{smilar}) and (\ref{su}) and choosing the following conditions,\begin{eqnarray}
  2D_{1}=-(4D_{7}+D_{8}), \textnormal{and}\ \ 2D_{1}=D_{9}, \label{suu} \end{eqnarray}
 $L_{1-11}$ takes the form,
\begin{eqnarray}
L_{1-11}=  D_{1}\bigg[A_{*}^{2}\left(K^{cd}K_{cd}-K^{2}\right)\mathcal{F}_{1}+4A_{*}^{2}\left(K K^{ab}-K^{a}_{d}K^{db}\right)F_{ab}+2 A_{*}^{2} h^{a b} \mathcal{L}_{n}  K_{a b}\mathcal{R}\nonumber\\-4A_{*}^{2}h^{bd}h^{ca}\mathcal{L}_{n}K_{cd}\mathcal{R}_{ab}-2\left(\mathcal{R}+K^{2}-K_{ab}K^{ab}\right) A_{*}K(\mathcal{L}_{n}A_{*})
\nonumber\\
+4A_{*}(\mathcal{L}_{n}A_{*})K^{a b}(\mathcal{R}_{ab}+KK_{ab}-K_{ac}K_{b}^{c})-2A_{*}^{2}{}_{\perp} R_{a b c d}K^{ac}K^{bd}+L(I)\bigg].\label{finals}
\end{eqnarray}

It is also worth mentioning here  that eq.(\ref{suu}) provide a relation $ 4D_{7}+D_{8}+D_{9}=0$, and from eq.(\ref{con2}) we find $D_{8}=D_{10}$. The terms with $D_{7}$ can not be removed in any case. 
%Further, for removing mixed term of the second derivative of the scalar field and the first derivative of the metric present in eq.(\ref{finals}). 
We  write the final part of the Lagrangian eq.(\ref{SE}) in this gauge  as,
\begin{eqnarray}
L_{12-16}=D_{15}\bigg[(\mathcal{L}_{n}A_{*}) A_{*}^{3} \left(- \tfrac{4}{3} K_{a}{}^{c} K^{ab} K_{bc} + 2 K \
K_{bc} K^{bc} -  \tfrac{2}{3} K^{3}\right)+ A_{*}^4 (K_{a}{}^{c} K^{ab} K_{b}{}^{d} K_{cd}\nonumber\\ -  \tfrac{4}{3} \
K K_{b}{}^{d} K^{bc} K_{cd} -  \tfrac{1}{2} K_{ab} K^{ab} \
K_{cd} K^{cd} + K^{2}K_{cd} K^{cd} -  \
\tfrac{1}{6} K^{4}) \bigg].
\end{eqnarray}
In order to ensure that our final action is free from mixed terms containing both the second derivative scalar field and the first derivative of metric, we choose $D_{15}=\frac{3D_{1}}{X}$, such that the total Lagrangian becomes,
\begin{eqnarray}
 L_{1-16}= D_{1}\bigg[A_{*}^{2}\left(K^{cd}K_{cd}-K^{2}\right)\mathcal{F}_{1}+4A_{*}^{2}\left(K K^{ab}-K^{a}_{d}K^{db}\right)F_{ab}+2 A_{*}^{2}  \mathcal{L}_{n}  K^{a b}\left(h_{ab}\mathcal{R}-2\mathcal{R}_{ab}\right)\nonumber\\-2\left(h_{ab}\mathcal{R}-2\mathcal{R}_{ab}\right)K^{ab} A_{*}(\mathcal{L}_{n}A_{*})+ A_{*}^2 (K_{a}{}^{c} K^{ab} K_{b}{}^{d} K_{cd}\nonumber\\ -  \tfrac{4}{3} \
K K_{b}{}^{d} K^{bc} K_{cd} -  \tfrac{1}{2} K_{ab} K^{ab} \
K_{cd} K^{cd} + K^{2}K_{cd} K^{cd} -  \
\tfrac{1}{6} K^{4})\bigg]+L(I). \label{fg}
\end{eqnarray}

As we can see from eq.({\ref{fg}}), all quadratic double derivative terms and mixed terms containing both double derivatives of scalar field and single derivative of metric are missing from the total Lagrangian $L_{1-16}$. Now, the remaining problematic terms in $L_{1-16}$ are linear in double derivative of both scalar field and metric. Next, we discuss a possible way to construct a healthy theory in the presence of higher derivatives for our Lagrangian $L_{1-16}$ as in eq.({\ref{fg}}).

\section{Analysis for  Ghost Free Theory}
In this section, we focus on the remaining higher derivative terms present in eq.(\ref{fg}). At  the action level, these higher derivative  terms in eq.(\ref{fg}) can be written as,
\begin{eqnarray}
    S_{HD}=\int d^4x N \sqrt{-h}\ 2D_{1}(\phi,X)\frac{1}{N}\left[ A_{*}^{2} \dot{ K^{a b}}-(\dot{A_{*}}) K^{ab} A_{*}\right]\left(h_{ab}\mathcal{R}-2\mathcal{R}_{ab}\right).\label{sde}
\end{eqnarray}
Ref.\cite{Langlois:2015cwa} shows a methodology to analyze such higher derivative terms by looking at the degeneracy structure of the theory. First, we briefly recall the treatment described in \cite{Langlois:2015cwa}. They consider an action,
\begin{eqnarray}
S_{L}= \int d^{4}x \sqrt{-g} (f(\phi)R+ C^{\mu \nu ,\rho \sigma} \nabla_{\mu} \nabla_{\nu} \phi \nabla_{\rho} \nabla_{\sigma} \phi).\label{langlosis}
\end{eqnarray} 
 where	$  C^{\mu \nu ,\rho \sigma} $  satisfies the following symmetry condition,          
\begin{eqnarray}
  C^{\mu \nu ,\rho \sigma} =  C^{ \nu \mu ,\rho \sigma} = C^{\mu \nu , \sigma \rho} =  C^{ \rho \sigma , \mu \nu}.         
\end{eqnarray} 
With this property, its  general form  can be expressed as, 
  \begin{eqnarray}
   C^{\mu \nu ,\rho \sigma} = \frac{1}{2} \alpha_{1}(g^{\mu \rho} g^{\nu \sigma} +g^{\mu \sigma} g^{\nu \rho})+\alpha_{2} g^{\mu
  \nu} g^{\rho \sigma} +\frac{1}{2} \alpha_{3} ( \phi^{\mu} \phi^{\nu}  g^{\rho \sigma}+ \phi^{\sigma} \phi^{\rho} g^{\mu \nu})\nonumber \\ +\frac{1}{4} \alpha_{4}( \phi^{\mu} \phi^{\rho}  g^{\nu \sigma}+ \phi^{\nu} \phi^{\sigma} g^{\mu \rho} + \phi^{\mu} \phi^{\sigma}  g^{\rho \nu}+ \phi^{\mu} \phi^{\sigma} g^{\mu \rho})+ \alpha_{5} \phi^{\mu} \phi^{\nu} \phi^{\rho} \phi^{\sigma},
  \end{eqnarray}
 where, $f(\phi)$ is only a function of $\phi$ and $\phi^{\mu}=\nabla^{\mu}\phi$. %Our notation is similar to that of ref. \cite{Langlois:2015cwa}.
 Using  eq.(\ref{R1}), we can write the kinetic term for action eq.(\ref{langlosis}) as,
    \begin{eqnarray}   L_{\phi}^{kin}=   C^{ab,cd} \lambda_{ab} \lambda_{cd} A_{*}^{2}+ 2 C^{ab,cd}  \Lambda_{ab}^{ef} \lambda_{cd} \dot{A}_{*} K_{ef} +   \Lambda_{ab}^{ef} \Lambda_{cd}^{gh} K_{ef} K_{gh}.
    \end{eqnarray}  
 The different coefficients of kinetic matrix in unitary gauge can be expressed as,
 \begin{eqnarray} 
 \begin{aligned}
 \mathcal{A}&=  C^{ab,cd} \lambda_{ab} \lambda_{cd}= \dfrac{1}{N^{2}} [\alpha_{2} + \alpha_{1} -(\alpha_{3}+ \alpha_{4}) A_{*}^{2} + \alpha_{5} A_{*}^{4}], \\&
  \mathcal{B}^{ef} = C^{ab,cd}  \Lambda_{ab}^{ef} \lambda_{cd} = \beta_{1} h^{ef},
\\&  \mathcal{K}^{ef,gh}=\Lambda_{ab}^{ef} \Lambda_{cd}^{gh}= \kappa_{1} h^{a(c} h^{d)b}+ \kappa_{2} h^{ab} h^{cd},
 \end{aligned}
\end{eqnarray}
   where,
  $ \beta_{1}=\frac{A_{*}}{2N}(2 \alpha_{2}-\alpha_{3} A^{2}_{*})$, $\kappa_{1}= \alpha_{1} A_{*}^{2} $, and $ \kappa_{2}= \alpha_{2}  A_{*}^{2} $. \\ 
Now, by using the identity $n^{a}\nabla_{a}\left(K_{bc}\right)=\frac{1}{N}\dot{K^{ab}}-\frac{N^{i}}{N}\mathcal{D}_{i}K^{ab}$ (ignoring the spatial derivatives terms), and $X=-A_{*}^2$, eq.(\ref{sde}) becomes,
\begin{eqnarray}
    S_{HD}=\int d^4x N \sqrt{-h}\ 2D_{1}(\phi,-A_{*}^2)\left[ A_{*}^{2}\ n^{e}\nabla_{e}K^{a b}-( n^{e}\nabla_{e}{A_{*}}) K^{ab} A_{*}\right]\left(h_{ab}\mathcal{R}-2\mathcal{R}_{ab}\right),
\end{eqnarray}
 Integrating by parts, it yields,
 \begin{eqnarray}
   \begin{aligned}
    S_{HD}=&\int d^4xN \sqrt{-h} \bigg[- D_{1} n^{e}\nabla_{e}(A_{*}^{2})-A_{*}^{2} n^{e}\nabla_{e}( D_{1})- D_{1} A_{*}^{2}K -D_{1}( n^{e}\nabla_{e}{A_{*}}) A_{*}\bigg]K^{a b}\left(h_{ab}\mathcal{R}-2\mathcal{R}_{ab}\right)\\&-\int d^4xN \sqrt{-h}\  D_{1}K^{ab }n^{e}\nabla_{e}\left(h_{ab}\mathcal{R}-2\mathcal{R}_{ab}\right).\label{fgd1}
      \end{aligned}
\end{eqnarray}
To simply further, we use the following relation,
 \begin{eqnarray}
    n^{e}\left(\nabla_{e}D_{1}(\phi,-A_{*}^2)\right)=\frac{1}{N}\left(-\frac{\partial(D_{1}(\phi,-A_{*}^2)}{\partial (-A_{*}^2)}2A_{*}\dot{A}_{*}+N\frac{\partial(D_{1}(\phi,X)}{\partial \phi}A_{*}\right)-\frac{N^{i}}{N}\mathcal{D}_{i}D_{1}. \label{rela}
    \end{eqnarray}
  By inserting eq.(\ref{rela}) into eq.(\ref{fgd1}), we obtain,
   \begin{eqnarray}
       \begin{aligned}
    S_{HD}&=\int d^4xN \sqrt{-h} \bigg[ \frac{A_{*}}{N}\left(-3D_{1}+2A_{*}^{2}\frac{\partial(D_{1}(\phi,-A_{*}^2)}{\partial (-A_{*}^{2})}\right)\dot{A}_{*}K^{ab}\left(h_{ab}\mathcal{R}-2\mathcal{R}_{ab}\right)-D_{1} A_{*}^{2}KK^{ab}\left(h_{ab}\mathcal{R}-2\mathcal{R}_{ab}\right)
    \\&+\bigg\{-\frac{\partial(D_{1}(\phi,-A_{*}^2)}{\partial \phi}A_{*}^3+N^{i}A_{*}^{2}\mathcal{D}_{i}D_{1}+\frac{N^{i}}{N}D_{1}\mathcal{D}_{i}(A_{*}^{2})\bigg\} K^{ab}\left(h_{ab}\mathcal{R}-2\mathcal{R}_{ab}\right)\bigg]\\&-\int d^4xN \sqrt{-h}\  D_{1}K^{ab}n^{e}\nabla_{e}\left(h_{ab}\mathcal{R}-2\mathcal{R}_{ab}\right).
      \end{aligned}
\end{eqnarray}
Kinetic part of the corresponding Lagrangian is,
\begin{eqnarray} 
L_{HD}^{kin}=  2 \mathcal{B}^{ab}_{1-16} \dot{A}_{*} K_{ab} + \mathcal{K}_{1-16}^{ab,cd} K_{ab} K_{cd},
    \end{eqnarray}  
where,
\begin{eqnarray}
\begin{aligned}
\mathcal{B}^{ab}_{1-16}&=\frac{1}{N}A_{*}\left(-3D_{1}+2A_{*}^{2}\frac{\partial(D_{1}(\phi,-A_{*}^{2})}{\partial (-A_{*}^{2})}\right)( h^{ab}\mathcal{R}-\mathcal{R}^{ab})&\\
\mathcal{K}_{1-16}^{ab,cd}&=-\frac{1}{N}D_{1}h^{cd} A_{*}^{2}\left(h^{ab}\mathcal{R}-2\mathcal{R}^{ab}\right).
\end{aligned}
\end{eqnarray}
 Adding both $L_{HD}^{kin}$ and $L_{\phi}^{kin}$ (kinetic parts of $S$ and $S_{HD}$ respectively), we get
 \begin{eqnarray}
 \begin{aligned}
     L^{kin}= L_{\phi}^{kin}+L_{HD}^{kin} =\mathcal{A}A_{*}^{2}+ 2 \tilde{B}^{ab} \dot{A}_{*} K_{ab} + \tilde{\mathcal{K}}^{ab,cd} K_{ab} K_{cd},
     \end{aligned}
 \end{eqnarray}
where, 
\begin{eqnarray} 
\begin{aligned}
\tilde{B}^{ab} =& \mathcal{B}^{ab}+ \mathcal{B}^{ab}_{1-16}, 
&\\ \tilde{\mathcal{K}}^{ab,cd}=&\mathcal{K}^{ab,cd}+\mathcal{K}^{ab,cd}_{1-16}.
\end{aligned}
\end{eqnarray}
Now the kinetic matrix becomes,
  \begin{eqnarray} 
  \mathcal{M}= \begin{bmatrix}
   \mathcal{A} & \tilde{B}^{cd} \\
  {\tilde{B}^{ab}} &  \tilde{K}^{ab,cd}
  \end{bmatrix}. 
  \end{eqnarray}
For this matrix, we find two possibilities for degeneracy conditions.
 \begin{enumerate}
\item Case I: The forms of $\mathcal{B}^{ab}_{1-16}$ and  $ \mathcal{K}_{1-16}^{ab, cd}$ indicate that for any metric whose intrinsic curvature of Ricci scalar $\mathcal{R}$ and  Ricci tensor $\mathcal{R}_{ij}$ is zero, we do not expect any contribution from higher derivative terms. Hence the class of theories with such metrics are free from the Ostrogradsky ghost. As an example, we will study the flat FRW metric in the next section.

\item  Case II:   The condition $\left(-3D_{1}+2A_{*}^{2}\frac{\partial(D_{1}(\phi,-A_{*}^{2})}{\partial (-A_{*}^{2})}\right)=0$ implies no contribution to kinetic matrix from higher derivative terms. Solution to the above condition is $D_{1}=\frac{\mathcal{C}}{A_{*}^{3}}.$ For this case, eq.(\ref{sde}) reduces to,
\begin{eqnarray}
    S_{HD}=\int d^4x N \sqrt{-h}\left[\ -2\frac{1}{N}\dfrac{d}{dt}\left(\frac{K^{ab}}{A_{*}}\right)\left(h_{ab}\mathcal{R}-2\mathcal{R}_{ab}\right)\right],\label{sde1}
\end{eqnarray}
Further, using the relation $n^{a}\nabla_{a}\left(\frac{K_{bc}}{A_{*}}\right)=\frac{1}{N}\dfrac{d}{dt}\left(\frac{K^{ab}}{A_{*}}\right)-\frac{N^{i}}{N}\mathcal{D}_{i}\left(\frac{K^{ab}}{A_{*}}\right)$, we obtain
\begin{eqnarray}
    S_{HD}=\int d^4x N \sqrt{-h}\left[-2 \bigg\{n^{a}\nabla_{a}\left(\frac{K_{bc}}{A_{*}}\right)+\frac{N^{i}}{N}\mathcal{D}_{i}\left(\frac{K^{ab}}{A_{*}}\right)\bigg\} \left(h_{ab}\mathcal{R}-2\mathcal{R}_{ab}\right)\right].
\end{eqnarray}
which can be further simplified to,
\begin{eqnarray}
    S_{HD}=-\int d^4x N \sqrt{-h}\left[\ -\left(\frac{K^{ab}}{A_{*}}\right)n^{a}\nabla_{a}\left(h_{ab}\mathcal{R}-2\mathcal{R}_{ab}\right)\right]-\int d^4x N\sqrt{-h} K\frac{K^{ab}}{A_{*}}\left(h_{ab}\mathcal{R}-2\mathcal{R}_{ab}\right).\label{fisa}
\end{eqnarray}
Here, we have used the relation $\nabla_{a}n^{a}=K$. It is clear from eq.(\ref{fisa}) that our Lagrangian does not contain any higher derivative terms and as a result we can show that our theory is free from Ostrogradsky instabilities in the unitary gauge.
\end{enumerate}
\subsection{Working Examples}
Here, we first calculate the equation of motion for the flat FRW metric. In this case, both intrinsic curvature $\mathcal{R}$ and $\mathcal{R}_{ij}$ vanish which makes our Lagrangian free from higher derivative terms. However, we will see that FRW equations are modified and depend on the higher derivative terms nontrivially.\\
The FRW metric with lapse function $N(t)$ is given by,
\begin{eqnarray}
  ds^{2}=- N(t)^{2} dt^2+a(t)^2 (dx^2 +dy^2 +dz^2),
  \label{frw}
 \end{eqnarray} 
 where, a(t) is the scale factor. 
 %Then the Lagrangian $L_{1-16}$ in eq.(\ref{fg}) becomes,
%\begin{eqnarray}
 %L_{1-16}=(-36D_{7}-12D_{9})\,{\frac {  \dot{a}^{4}\ \dot{\phi}^{2}\ }{ a^{4} \ N^{6}} }.\label{final}
%\end{eqnarray}
%For further calculation, we assume $D^{'}=-36D_{7}-12D_{9}\neq0.$
Here, our action is,
\begin{eqnarray}
  S= \int d^{4}x \sqrt{-g} \left[\frac{R}{2\kappa}-\frac{1}{2}g^{\mu\nu}\nabla_{\mu}\phi\nabla_{\nu}\phi-V(\phi)+L_{1-16}\right],\
  \label{general}
\end{eqnarray}
where $\kappa^{2}=\frac{1}{M_{p}^{2}}$, $M_{p}$ is the four-dimensional Planck mass, and $V (\phi)$ is the scalar potential. 
%We add the Lagrangian $ L_{1-16}$  along with  usual Einstein Hilbert(E-H) and scalar field Lagrangian. 
For the metric eq.(\ref{frw}), the action becomes,
\begin{equation}
  S_{FRW}= \int d^{4}x a^{3}N \left[\frac{\ddot{a}-\dot{a}^2}{a^{2}N^{2}}+\frac{\dot{\phi}^{2}}{2N^2}-V(\phi)+D^{'}(\phi,X)\,{\frac {\dot{a}^{4} \dot{\phi}^{2}}{ a^{4} \ N^{6}} }\right],\end{equation}
  where we have defined $D^{'}=-36D_{7}-12D_{9}\neq0.$
Then, the equations of motion take the form,
\begin{eqnarray}
\begin{aligned}
-9 D^{'} H^{4}\dot{\phi}^{2}+ H^{3}\left(8\frac{\partial D^{'}}{\partial X}
\dot{\phi}^{3}\ddot{\phi}-4\frac{\partial D^{'}}{\partial \phi}  \dot{\phi}^{3}-8 \,D^{'}
\dot{\phi}\ddot{\phi} \right) -12\,D^{'} \dot{\phi}^{2} H^{2} \dot{H}&\\ +3\left(\frac{1}{2}\dot{\phi}^{2}-\,V \left( \phi \right)\right)+
 9\frac{H^{2}}{\kappa}+6\frac{\dot{H}}{\kappa} &=0,
&\\
 -\frac{1}{2}(V \left( \phi \right) +\dot{\phi}^{2})+3\, \frac{H^{2}}{\kappa}+ H^{4}\left(2\,\frac{\partial D^{'}}{\partial X} 
\dot{\phi}^{4}-5D^{'}
\dot{\phi}^{2}\right)&=0,
&\\
 H^{5} \left(-\,6 \frac{\partial D^{'}}{\partial X} \dot{\phi}^{3}+6 D^{'}\dot{\phi}\right)+\,H^{4}\left( 4 \frac{\partial^2 D^{'}}{\partial X^{2}}\dot{\phi}^4\ddot{\phi}-2 \frac{\partial^2 D^{'}}{\partial X \partial \phi}\dot{\phi}^4
-10\frac{\partial D^{'}}{\partial X}\dot{\phi}^{2}\ddot{\phi}+\frac{\partial D^{'}}{\partial \phi}\dot{\phi}^{2}+2D^{'}\ddot{\phi}\right)&\\+H^{3}\left(-8\,\frac{\partial D^{'}}{\partial X} \dot{\phi}^{3}  \dot{H}+8D^{'}\dot{\phi}\dot{H}\right)-3\, \dot{\phi} H - \frac{\partial V   \left( \phi\right)}{\partial \phi} -\,\ddot{\phi} &=0.
\end{aligned}
\end{eqnarray}
Note that all the equations of motion are of second-order and FRW equations get modified in the presence of $L_{1-16}$. Such modification also considered by taking Gauss-Bonnet term in\cite{Oikonomou:2020tct,Ai:2020peo,Hwang:2005hb,Noh:2001ia}.
%However, the detailed cosmology of this theory will be discussed elsewhere.

We can further extend the analysis described above to closed and open FRW universes, which can be expressed by the metric,
 \begin{equation}
     ds^2=-dt^2+a(t)^{2}\left(\frac{dr^2}{1-\mathcal{K}r^2} +r^{2}d\theta^2+r^2sin^2\theta d\phi^{2}\right),
 \end{equation}
written in spherical polar coordinates $r$, $\theta$, $\phi$, where  $\mathcal{K} $ is the spatial curvature. It is clear that $(h_{ij}\mathcal{R} -2R_{ij}) \neq 0$ for the open (K=-1) and the closed (K=1) metric. This leaves our analysis of section IV of finding degeneracy condition irrelevant in these cases. Here, we simply consider the Lagrangian $L_{1-16}$ to show how degeneracy conditions $\left(D_{1}=\frac{1}{|X|^{\frac{3}{2}}}\right)$ remove the higher derivative contributions at the action level. 

For the form of $L_{1-16}$ in eq.(\ref{fg}), $S_{HD}$ becomes,
 \begin{eqnarray}
    S_{HD}=\int d^4x\ a^{3}\left[( -36D_{7}-12D_{9})\,{\frac { \left( \dot{a}+\mathcal{K} \right) ^{2} \dot{\phi}^2}{ a^{4}}}\right]+12\int d^4x \ \mathcal{K}\dfrac{d}{dt}\left(\dfrac{\dot{a}}{\dot{\phi}}\right).
 \end{eqnarray}
 Here, we have shown that no higher derivatives terms appear in action for this case, which acts as an example of degeneracy.
 
\section{Conclusion}

In the present work, we introduce a higher derivative scalar-tensor model in which both the double-time derivatives of metric and scalar field appear.  We have considered a most general curvature coupling (up to quadratic order) with a scalar field. 
There are 16 different terms in our Lagrangian which are given by,
 \begin{eqnarray}
\begin{aligned}
L_{1-16}&=\bigg(D_{1} R  g^{c e} g^{d f} +D_{2}R g^{c d} g^{e f}  
+D_{3} g^{a c} g^{b e} g^{d f} R_{a b} 
+D_{4} g^{a e} g^{c d} g^{b f} R_{a b}
+ D_{5}g^{c a} g^{d b} g^{l e} g^{m f} R_{a b l m} \\&+
 D_{6}g^{c a} g^{f b} g^{l e} g^{m d} R_{a b l m}\bigg)\nabla_{c} A_{d} \nabla_{e} A_{f}+\bigg(D_{7}g^{a b} g^{c d} g^{e f}  R_{a b}R_{c d}+
+ D_{8}g^{a b} g^{c e} g^{d f}  R_{a b}R_{c d}\\&+ D_{9}g^{a c} g^{b d} g^{e f}  R_{a b}R_{c d}+ D_{10}g^{a c} g^{b e} g^{d f}  R_{a b}R_{c d}+ D_{11}g^{c r} g^{q b} g^{d f} g^{s e} g^{a p} R_{a c b d}R_{p q r s}\bigg)A_{e} A_{f}
 \\&+\bigg(D_{12}g^{ab} g^{cd} g^{ef} g^{pq}+D_{13} g^{ae} g^{dp} g^{cf} g^{qb}+D_{14}g^{ac} g^{bd} g^{ep} g^{qf}+D_{15}g^{ab} g^{cd} g^{pf} g^{qe}\\&+D_{16}g^{ab} g^{cq} g^{pf} g^{de} \bigg) \nabla_{a} A_{b}\nabla_{c} A_{d} \nabla_{e} A_{f} \nabla_{p} A_{q}+\lambda^{a} (\nabla_{a} \phi-A_{a}.
\end{aligned}
\end{eqnarray}
While working with the general space-time metric, we cannot find any degeneracy present in our theory. However, in unitary gauge, by choosing the following relations among different coefficients,
\begin{eqnarray}
\begin{aligned}
  D_1&=-D_{2}= -\dfrac{D_3}{4}=\dfrac{D_4}{4}=\dfrac{D_{6}}{2},\\ D_{15}& =-6D_{12}= D_{13}=-2 D_{14}=-\frac{3}{4} D_{16},\\4D_{7}&+2D_{8}+D_{9}+D_{10}=0,\qquad D_{9}+D_{11}=0.
  \end{aligned}
  \end{eqnarray} 
  After employing the above relations, the remaining linear in double time derivative terms are removed by the conditions which are,
  \begin{eqnarray}
    2D_{1}=-(4D_{7}+D_{8}),\ \  2D_{1}=D_{9},\ \
    D_{15}=\frac{3D_{1}}{X}.
   \end{eqnarray}
   Finally, it results in working with only  $D_{1}.$  Further, we extend the analysis done in ref. \cite{Langlois:2015cwa} by adding the Lagrangian
   $L_{1-16}$ to their action.

  Our primary focus here is to check the validity of this theory in a unitary gauge. In this context, we also find under what condition these higher derivative terms do not appear in the kinetic matrix of total action $S_L+S_{HD}.$  Degeneracy conditions give rise to two choices where the first one applies to metric for which the combination $h^{ab}\mathcal{R}-\mathcal{R}^{ab} $ vanishes. The second one would be more interesting, which works for a choice of $D_1$ which is $D_{1}=\frac{\mathcal{C}}{A_{*}^{3}}.$
  
 Under these degeneracy conditions, we have shown that our theory is free from the Ostrogradsky ghost. This situation holds for a particular combination of 16 coefficients of different terms in Lagrangian $L_{1-16}$, which shows that our theory does not suffer from higher derivative terms in the unitary gauge. This can be considered as a theory analogous to the class of U-DHOST theories\cite{Gao:2014soa,Gao:2014fra,Fujita:2015ymn,Gao:2018znj,Gao:2019lpz,Gao:2018izs,Gao:2019liu,DeFelice:2018ewo, DeFelice:2021hps}.
   
   Finally, we derive the action for background flat, closed and open FRW metrics as specific examples. In this direction, we obtain a modified FRW equation for the flat case. The background cosmology of this model will be taken as a future project. We plan to conduct a similar analysis for the most general choice of time-independent symmetric and asymmetric space-time in the future. 
 \section{ACKNOWLEDGEMENT}
This work is partially funded by DST (Govt. of India), Grant No. SERB/PHY/2017041. Calculations are performed using xAct packages of Mathematica.

\end{document}